\documentclass[prd,preprint,nofootinbib,showpacs]{revtex4}
\usepackage{amsmath}

\newcommand{\bi}{\bar{\imath}}
\newcommand{\bj}{\bar{\jmath}}
\newcommand{\bk}{{\bar k}}

\renewcommand{\b}[1]{\bar{#1}}
\newcommand{\be}{\begin{equation}}
\newcommand{\ee}{\end{equation}}
\newcommand{\bea}{\begin{eqnarray}}
\newcommand{\eea}{\end{eqnarray}}

\newcommand{\Del}{\nabla}
\newcommand{\del}{\partial}

\newcommand{\N}{\mathcal{N}}
\renewcommand{\ap}{\alpha'}

\begin{document}
\preprint{hep-th/0307142}

\author{Andrew R. Frey}
\email{frey@vulcan.physics.ucsb.edu}
\affiliation{Department of Physics\\
University of California\\ 
Santa Barbara, CA 93106, USA}

\author{Mariana Gra{\~n}a}
\email{mariana@cpht.polytechnique.fr}
\affiliation{Centre de Physique Th{\'e}orique\\
Ecole Polytechnique\\
F-91128 Palaiseau Cedex, France}

\title{Type IIB Solutions with Interpolating Supersymmetries}


\begin{abstract}

We study type IIB supergravity solutions with four supersymmetries that 
interpolate
between two types widely considered in the literature: the dual of 
Becker and Becker's compactifications of M-theory to 3 dimensions 
and the dual of Strominger's
torsion compactifications of heterotic theory to 4 dimensions. 
We find that for all intermediate solutions the internal manifold is
not Calabi-Yau, but has $SU(3)$ 
holonomy in a connection with a torsion given by the
3-form flux.  All
3-form and 5-form fluxes, as well as the dilaton, 
depend on one function appearing in the 
supersymmetry spinor, which satisfies a nonlinear differential equation. 
We check
that the fields corresponding to a flat bound state of D3/D5-branes lie in our
class of solutions.  The relations among supergravity fields that we 
derive should be useful in studying new gravity duals of gauge theories,
as well as possibly compactifications.

\end{abstract}

\maketitle

\section{Introduction}\label{s:intro} 

Supersymmetric ${\mathcal N}=1$ warped solutions of IIB supergravity 
have played an
extensive role in gauge/gravity duality and string compactification.  
Unfortunately, the
general supersymmetric solution is not known. 
Recent work has involved two special cases with 
four dimensional Poincar{\'e} invariance, which can be characterized
by the
form of the ten-dimensional supersymmetry spinor 
$\varepsilon=(\varepsilon^1, \varepsilon^2)$.  These Majorana-Weyl 
spinors can be decomposed as
\bea
\varepsilon^1 &=& \zeta \otimes \chi_1 + \zeta^* \otimes \chi_1^* \nonumber\\
\varepsilon^2 &=& \zeta \otimes \chi_2 + \zeta^* \otimes \chi_2^*.
\label{decomp}
\eea
Here $\zeta$ is a four-dimensional chiral spinor, 
$\Gamma^{(4)} \zeta = \zeta$, and
$\chi_{1,2}$ are six-dimensional chiral spinors, $\Gamma^{(6)} 
\chi_i = -\chi_i$.
Each independent pair $(\chi_1,\chi_2)$ gives rise to one $D=4$ 
supersymmetry.
The two special cases are then
\begin{eqnarray}
\textnormal{Type A(ndy):}&& \chi_2 = 0  \nonumber\\
\textnormal{Type B(ecker):}&& \chi_2 = i\chi_1\ .
\end{eqnarray}
The behavior of the spinor correlates with that of the complex 
three-form flux
$G_{(3)}$.  In type A solutions there is only NS-NS 3-form flux, 
which means that $G_{(3)}$  must be imaginary.  In fact, only the NS-NS
background is nontrivial.
In type B
solutions, the 5-form is non-vanishing.  Also, 
$G_{(3)}$  must be imaginary self-dual; more specifically, 
it must be of $(2,1)$ and primitive with respect to the complex
structure of the
transverse space. Vanishing $G_{(3)}$ also gives type B solutions.
Pure brane systems are of one or the other of these types: the 
NS5-brane is of type A (and the D5-brane is of the S-dual type), 
and the D3-brane and D7-brane are of type B.

The type A solutions are closely related to the warped heterotic 
solutions
found by Strominger \cite{Strominger:1986uh}.  (Among other papers,
the IIB supersymmetry conditions for type A solutions were discussed in detail in
\cite{Ivanov:2000fg,Buchel:2001qi}.) The Maldacena-Nu{\~n}ez solution 
\cite{Maldacena:2000yy} is a notable
AdS/CFT example of this type.  Conditions to have $\N=2$
supersymmetry in this class of solutions are discussed in  
\cite{Gauntlett:2003cy}, and an $\N=2$ AdS/CFT example is
\cite{Gauntlett:2001ps}.
Type A compactifications
have been reconsidered recently in 
\cite{Becker:2002sx,Kachru:2002sk,Becker:2003yv}, because they are related
to type B compactifications by a series of U-dualities.

The type B solutions are dual to M theory solutions found by Becker and
Becker \cite{Becker:1996gj,Gukov:1999ya}.  
In the M theory form, the corresponding
restriction on the supersymmetry spinor is that it have definite
eight-dimensional chirality.
The explicit IIB form was obtained in
\cite{Grana:2000jj,Gubser:2000vg} for the special case of a constant 
dilaton-axion, 
and in \cite{Kehagias:1998gn,Grana:2001xn}
for nonconstant dilaton-axion.  Such
solutions
have played an important role in gauge/gravity duality.  
Along with the standard $AdS_5\times S^5$ solution, the $\N = 1$ 
conifold
fractional brane solution \cite{Klebanov:2000hb} is of this form, 
as well as its $\N = 2$
generalization \cite{Bertolini:2000dk,Polchinski:2000mx}.
Type B compactifications have been the focus of intense
interest; 
simple compact manifolds were studied in 
\cite{Dasgupta:1999ss,Verlinde:1999fy,Greene:2000gh,Kachru:2002he,
Frey:2002hf,Tripathy:2002qw} and general cases were examined in
\cite{Chan:2000ms,Giddings:2001yu}, including supersymmetry breaking
cases. Giddings, Kachru and Polchinski \cite{Giddings:2001yu} showed
that 
compact
solutions involving D3-branes, O3 planes, D5-branes wrapped on
collapsed 2-cycles and D7 branes wrapped on $K3$ are all of type B
form. 
Notably, \cite{Kachru:2003aw} has constructed a de Sitter solution
of string theory by adding nonperturbative effects to type B 
compactifications.

In this paper, we will describe a class of solutions that interpolates
between type A and B solutions (effectively tracing out a curve in 
spinor
space), studying the fermion supersymmetry transformations 
directly.  Our solutions would then correspond to D3/NS5 bound states
and eventually D7-branes when we include a nontrivial axion,
where in one extrema (type A) the D3 (and D7)-charge vanishes, while in the
other (type B) the 5-brane wraps a vanishing 2-cycle. 
We found it easiest to describe solutions that interpolate
between the S-dual of type A solutions, which we 
will
call type C (for convenience) 
and correspond to D5-branes with spinors $\chi_2=-\chi_1$, 
and the type B solutions, since for both
of them the 
spinors $\chi_1$ and $\chi_2$ have the same norm.  We can then 
S-dualize
to find solutions that interpolate between the NS5-brane type A and 
type B solutions.  We start in the next section by reviewing 
characteristics
of type A, B, and C solutions; then in section \ref{s:interp}, we
describe the actual interpolating solution, keeping a trivial R-R scalar
for simplicity.  
In section \ref{s:generalize}, we explain how to include a nontrivial
R-R scalar (as sourced by D7-branes) and also how to generalize the
type of 5-brane charge (by duality transformations).
Because bound states of 
branes
should correspond to some intermediate type of supersymmetry solution,
we examine the supergravity solution of a D3/D5-brane bound state 
\cite{Breckenridge:1997tt,Costa:1998zd} in
section \ref{s:bound} and show that it is in the class described in
\ref{s:interp}. Finally, we discuss our results in section \ref{s:discuss}.

\section{Known Solutions of IIB Supergravity}\label{s:known}

All the solutions we study preserve $SO(3,1)$ invariance, 
so we start with the most general
ansatz for the background metric and 5-form flux that preserve such 
invariance:
\bea
ds^2=e ^{2A}\eta_{\mu \nu} dx^{\mu}dx^{\nu} + ds_6^2 \nonumber\\
F_{\mu\nu\lambda\rho m}= e^{-4A} \epsilon_{\mu\nu\lambda\rho}\partial_m 
h\ . \label{ansatz}
\eea
The 3-forms are completely on the transverse space; we allow branes 
that
are extended in the four $x^\mu$ directions.
Greek subindices take values from 0 to 3, and indicate directions of 
the 
$SO(3,1)$ symmetry, 
while Roman subindices take values from 4 to 9. 
$\epsilon_{\mu\nu\lambda\rho}$ has coordinate  
indices, i.e. $\epsilon_{0123}= e^{4A}$. $A$ and $h$ 
in (\ref{ansatz}) are functions of $x^m$.

In the following two subsections, we will write the known type A, C and B 
solutions. We will not show  how to get these solutions from
supersymmetry constrains, since that will become clear when working
out the interpolating solutions. 

\subsection{Types A and C} \label{ss:typesAandC}

The type A solution is the dual of Strominger's heterotic ``superstrings
with torsion'' \cite{Strominger:1986uh}. The IIB version of it is
the supergravity solution describing NS5 branes wrapped on 2-cycles.
We work in the string frame, in which the metric
is unwarped, $A=0$ with the internal metric 
Hermitean on a
complex manifold. Such a manifold has torsion, i.e.  $SU(3)$
invariant tensors are parallel in a connection given by  
the Levi Civita connection
plus a torsion. This torsion is 
nothing else than
the NS-NS 3-form flux, and it is related to the complex structure of
the manifold by
\be
H= i ({\bar \partial} - \partial) J
\label{torsionJ}
\ee
The manifold is endowed with a holomorphic (3,0) form $\Omega$, whose norm
fixes the dilaton
\be
\phi=\phi_0 -\frac{1}{2} \ln |\Omega |
\ee
For Calabi-Yau manifolds $|\Omega |$ is constant, and $J$ is
closed. In this case, we have instead
\be
d^{\dagger} J+ i({\bar \partial} - \partial)\ln |\Omega | =0
\ee

We will see where these equations come from when working out the
interpolating solution. 

In the type C, or D5-brane solution, the form is much the same, with 
$H_{(3)}\to F_{(3)}$ and now $A=\phi/2$ and $ds_6^2 = e^{\phi} d\tilde s_6^2$
with $d\tilde s_6^2$ the complex manifold with torsion.  The solutions
corresponding to $(p,q)$5-branes are somewhat more complicated, as they
include a nontrivial R-R scalar.  We do not describe them here.

\subsection{Type B}\label{ss:typeB}

Type B solutions, corresponding to regular D3 branes, are of the 
``warped Calabi-Yau'' form:
\be
ds_6^2= e^{-2A} d\tilde{s}_6^2
\label{Bmetric}
\ee where $d\tilde{s}_6^2$ is a metric on a Calabi-Yau manifold.  The
square of the warp factor, usually called $Z$, obeys a Poisson equation:
\be -\tilde{\nabla}^2 (e^{-4A})=(2 \pi)^4 g \alpha'^2 \rho_3  \ee
where $\rho_3$ is the density of D3-branes.  There is 5-form flux of
the form stated in the ansatz (\ref{ansatz}),  where $h$ is \be
h=\frac{1}{g} e^{4A}
\label{Bh}
\ee and the dilaton is constant $e^{\phi}=e^{\phi_0}=g$.  (In fact,
the entire complex dilaton-axion $\tau=C+ie^{-\phi}$ is constant.)

We can add 3-form flux to this solution, which can be sourced by
fractional branes (D5-branes wrapped  on collapsed 2-cycles). The
condition that we get from supersymmetry is that the combination
$G_{(3)}=F_{(3)}-\tau H_{(3)}$ should be $(2,1)$ and primitive with
respect to the complex structure of the Calabi Yau space, which means
\be G_{ijk} = G_{ij}{}^j = G_{\bar\imath\bar\jmath\bar k} =
G_{\bar\imath\bar\jmath k} = 0\ . \label{21}
\end{equation}

This 3-form flux acts like a source in the Poisson equation for the 
warp factor, which gets modifies to
\be 
-\tilde{\nabla}^2 (e^{-4A})=(2 \pi)^4 g \alpha'^2 \tilde{*}_6\rho_3+ g^2
 G_{pqr} 
G^{\widetilde{pqr}*}\ ,
\label{warppoisson}\ee
where $\rho_3$ is the 6-form number density of D3-brane charge.

It is also known how to include a nontrivial axion-dilaton, as in F-theory,
in the presence of D7-branes \cite{Kehagias:1998gn,Grana:2001xn}.  The
axion-dilaton is holomorphic $\b\del\tau=0$, and the five-form now
satisfies $h=e^{4A-\phi}$.  We will see how this comes about below.

\section{Interpolating solution}\label{s:interp}

In this section we will show how to get interpolating solutions 
between 
type B and type C solutions.

The type IIB supersymmetry variations in the string frame for bosonic
backgrounds are \cite{Hassan:1999bv,Bergshoeff:1999bx}
\begin{eqnarray}
\delta\lambda&=& \frac{1}{2} \Gamma^M \partial_M \phi \varepsilon  
-\frac{e^\phi}{2}
\Gamma^{M}F_{M}\, (i\sigma^2)\, \varepsilon - \frac{e^\phi}{24}
\Gamma^{MNP}F'_{MNP}\, \sigma^1\, \varepsilon 
-\frac{1}{24}\Gamma^{MNP}H_{MNP}
\, \sigma^3\varepsilon \label{deltalambda}\\
\delta\psi_{M}&=&  \Del_{M} \varepsilon +\frac{1}{8} e^{\phi}\,\Gamma^N 
\Gamma_M\,
F_{N} (i \sigma ^2) \varepsilon -\frac{1}{8}
\Gamma^{PQ} H_{MPQ}\, \sigma^3 \varepsilon  \nonumber\\
&&+ \frac{1}{48} e^{\phi}\,
\Gamma^{PQR} \Gamma_M \, F'_{PQR}\, \sigma^1 \varepsilon 
 + \frac{1}{16 \cdot \,5!} e^{\phi}\,
\Gamma^{PQRST} F_{PQRST}\, \Gamma_M\, (i\sigma^2) \varepsilon\ .
\end{eqnarray}
As before, 
$\varepsilon=(\varepsilon^1,\varepsilon^2)$ are two Majorana-Weyl
spinors of negative 10d chirality, and $\sigma^i$ are the Pauli
matrices that act on the column vector $(\varepsilon^1,\varepsilon^2)$.  
We have defined $F'_{(3)}
=F_{(3)}-CH_{(3)}$.  
These variations vanish in supersymmetric solutions.  We use the following
decomposition for the gamma matrices:
\be\label{gammadecomp}
\Gamma^\mu = \gamma^\mu\otimes 1\ ,\ \Gamma^m = \Gamma^{(4)}\otimes
\gamma^m \ee
with $\Gamma^{(4)}= (i/4!) \epsilon_{\mu\nu\lambda\rho}
\gamma^{\mu\nu\lambda\rho}$.

\subsection{Solving Supersymmetry Conditions}\label{ss:interp1}

For convenience, we will set the axion  $C$ to zero and reintroduce it later. 
A constant nonzero $C$ trivially replaces $F\to F'$ in the below results.
From $\delta \psi_{\mu}=0$, we get
\be
\gamma^{\mu}\Gamma^{(4)} \left(\frac{1}{2} \gamma^n \partial_n A  +\frac{i}{8}
e^{-4A} e^{\phi} \gamma^n \partial_n h \Gamma^{(4)} i\sigma^2 
-\frac{1}{48}e^{\phi} F \sigma^1 \right) \varepsilon=0
\ee
where $F\equiv F_{mnp} \gamma^{mnp}$. This means
\be
\left(\frac{1}{2} \gamma^n \partial_n A  +\frac{i}{8}
e^{-4A} e^{\phi} \gamma^n \partial_n h \Gamma^{(4)} i\sigma^2
-\frac{1}{48}e^{\phi} F \sigma^1\right)
 \varepsilon=0
\label{gravitino0}
\ee
From  $\delta \psi_{m}$ we get
\be
\Del_m \varepsilon -\frac{1}{8}H_m \sigma^3 \varepsilon + \frac{1}{8} e^{\phi} 
F_m \sigma^1 \varepsilon \\
-\frac{1}{48} e^{\phi} \gamma_m F \sigma^1 \varepsilon +\frac{i}{8}
e^{-4A} e^{\phi}\gamma^n \gamma_m \partial_n h \Gamma^{(4)} i\sigma^2 
\varepsilon =0
\ee
using (\ref{gravitino0}), the term involving $F$ can be written in terms 
of derivatives of the warp factor and 4-form flux. 
Using $\gamma^n \gamma_m=2\gamma^n\,_m+\gamma_m 
\gamma^n$ in the last term, everything combines to give
\be
\Del_m\epsilon -\frac{1}{2} \partial_m A - \frac{1}{2} \gamma_m\,^n 
\partial_n A \epsilon - 
\frac{i}{4}e^{-4A} e^{\phi}  \gamma_m{}^n \partial_n h \Gamma^{(4)} i 
\sigma^2 \epsilon \\
-\frac{1}{8} H_m \sigma^3 \epsilon + \frac{1}{8} F_m \sigma^1 
\epsilon=0
\label{gravitino1}
\ee
where $F_m\equiv F_{mnp} \gamma^{np}$ and similarly for $H_m$. 

The term that does not contain gamma matrices can be canceled by 
the term $\partial_m \varepsilon$
if we 
define $\varepsilon=e^{A/2} \varepsilon'$. We get for $\varepsilon'$
\be
\Del_m\epsilon' - \frac{1}{2} \gamma_m{}^n \partial_n A \varepsilon' -
\frac{i}{4} e^{-4A} e^{\phi}\gamma_m{}^n \partial_n h \Gamma^{(4)} i 
\sigma^2 \varepsilon' \\
-\frac{1}{8} H_m \sigma^3 \epsilon' + \frac{1}{8} F_m \sigma^1 
\varepsilon'=0
\label{gravitino2}
\ee

To go any further, we should make an ansatz for the spinors, 
and this is the point where 
general (Poincar{\'e} invariant) solutions become particular ones. 
The ansatz that we make
is the simplest one interpolating between type B and type C spinors:    
\bea
\varepsilon^{1'} = \zeta\otimes \chi &+& \zeta^* \otimes \chi^* \nonumber\\
\varepsilon^{2'} = i e ^{i\alpha} \zeta\otimes \chi &-& i  e ^{-i\alpha} 
\zeta^* \otimes \chi^* .
\label{ansatzspinors}
\eea
Following the notation of the introduction, $\chi_1=e^{A/2}\chi, 
\chi_2=ie^{i\alpha}\chi_1$.
Type B corresponds to $\alpha=0$ and type C to $\alpha=\frac{\pi}{2}$.  We 
will
let $\alpha$ vary over the compact manifold, which, as we will see, is indeed 
necessary to get solutions other than B and C.

Inserting this ansatz in (\ref{gravitino2}) and using
(\ref{relhA}) below\footnote{We don't need to use (\ref{relhA}) here, 
but 
it makes things look a little simpler.},  we get
\bea
\left(\Del_m +\frac{1}{2} e ^{2i \alpha} \gamma_m\,^n \partial_n A   
-\frac{1}{8} H_m  + \frac{i}{8} e ^{i \alpha} e^{\phi} F_m\right)
\chi=0 \label{grav1}\\
\left(\Del_m +\frac{1}{2} e ^{-2i \alpha}  \gamma_m\,^n \partial_n A   
+\frac{1}{8} H_m  - \frac{i}{8} e ^{-i \alpha} e^{\phi} F_m
+i\del_m\alpha\right) \chi=0
\label{grav2}
\eea
Now add up (\ref{grav1},\ref{grav2}) to get
\be
\left(\Del_m +\frac{1}{2}\cos(2\alpha)\del_n A\gamma_m{}^n
-\frac{1}{8} \sin\, \alpha e ^{\phi} F_m +\frac{i}{2}\del_m\alpha
\right) \chi =0\ .
\label{grav3}
\ee
Again, the term that does not involve gamma matrices can be canceled by taking $\chi= e^{-i\alpha/2}\chi^\prime$.
If we assume for now that 
\be
\cos(2\alpha)\del_m A = \del_m A^\prime
\label{AA'} 
\ee
is a 
total derivative, then we can rescale the internal metric
\be\label{rescale}
ds^2_6 = e^{-2A^\prime} d\tilde{s}_6^2\ \ee
to eliminate the warp factor terms,
which clearly agrees with type B at $\alpha=0$ 
($A=A'$ in type B, as can be seen from (\ref{Bmetric})).
Such a function $A'$ does not logically have to exist.  In fact, we can 
distinguish three (nonexclusive) cases: first, that $A'$ does exist; second,
that the manifold is complex; and, third, that $A'$ does not exist and
the internal manifold is not complex.  The calculations presented in this
section demonstrate that the first two cases are equivalent, but we do not
address the possibility of the third case.

Then we have   
\be\label{cov+torsion}
\left(\tilde{\Del}_m -\frac{1}{8} \sin\alpha\, e ^{\phi+2A^\prime} 
F_{mnp}
\tilde{\gamma}^{np}\right) \chi^\prime =0\ ,\ee
This means that there is a normalized spinor that is constant with respect to
a connection equal to the Levi-Civita connection $\tilde{\Del}_m$ 
plus a torsion 
$-(1/2) \sin\alpha\, e
^{\phi+2A^\prime} F_m$. 
This suffices to show that the manifold is complex (but in general not
K{\"a}hler). The
complex structure, being built out of the spinor, will also be 
covariantly constant with respect to this connection with torsion. 
We will comment on this later.  For now, we just need that the manifold
is complex and that we can take $\gamma^{\bi}\chi=0$ for our negative
chirality spinor.

The relationship between $h$ and the warp factor
that we used to get Eqs (\ref{grav1},\ref{grav2}), can be obtained from 
(\ref{gravitino0}) using our particular ansatz for
the spinors. By doing so, we get
\bea
\left(\frac{1}{2} \gamma^n \partial_n A  -\frac{1}{8}e^{-4A} e^{\phi} e 
^{i\alpha}
 \gamma^n \partial_n h 
-\frac{i}{48} e ^{i\alpha} e^{\phi} F \right)
 \chi=0 \nonumber\\
\left(\frac{1}{2} i e ^{i\alpha} \gamma^n \partial_n A  -\frac{i}{8} 
e^{-4A} e^{\phi}
 \gamma^n \partial_n h 
-\frac{1}{48} e^{\phi} F \right)
 \chi=0 \label{gravitino01}
\eea
along with the complex conjugate equations.
Multiplying the second by $-ie ^{i\alpha} $ and adding both up, we get
\be
e ^{i\alpha} e^{-4A} e^{\phi} \partial_m h= 2 (e ^{2i\alpha} +1) 
\partial_m A 
\ee
or
\be
\del_m h = 4e^{4A} e^{-\phi}\cos\alpha\del_m A
\label{relhA}
\ee
(which means the rhs is a total derivative, at least in patches).
We also see  
from (\ref{gravitino01}) that  
\be
F_{ijk}=F_{\bi\bj\bk}=0. 
\ee

The other equation that we get from the set of Eqs.(\ref{gravitino01}) is
\be\label{FA1}
e^\phi F_{ij}{}^j = 4\sin\alpha \del_i A\ .\ee
The complex conjugates of these equations follow from the $\chi^*$ 
eqns.

Now, let us turn to the two equations coming from the variation of the
dilatino (\ref{deltalambda}), $\delta \lambda=0$:
\bea
\frac{1}{2}\gamma ^m \partial_m \phi \chi -\frac{1}{24}H \chi
-\frac{i}{24} e ^{i\alpha} e ^{\phi} F \chi =0 \label{dilatino1} \\
 \frac{1}{2}\gamma ^m \partial_m \phi \chi + \frac{1}{24}H \chi
+\frac{i}{24}  e ^{-i\alpha} e ^{\phi} F \chi =0 \label{dilatino2} 
\eea
First of all, knowing that there is no (3,0) R-R 3-form flux, we can see that
there should not be (3,0) (or (0,3)) NS-NS flux. So the 3-form flux
$G_{(3)}$ is only (2,1) and (1,2),
as happens in both type B and C solutions.  

Now, substracting (\ref{dilatino2}) from (\ref{dilatino1}), we get
\be
H \chi + i \cos\alpha\, e ^{\phi} F \chi =0
\ee
which means
\be
\cos\alpha\, F_{ij}{}^j -i e ^{-\phi} H_{ij}{}^j =0\, .
\label{tracesFH2}
\ee
For type B, this is the
primitivity condition for $G_{(3)}$, $G_{ij}\,^j=0$. 

Finally, adding  (\ref{dilatino2}) and (\ref{dilatino1}), we get the 
equation 
\be
\frac{1}{2}\gamma ^m \partial_m \phi \chi +
\frac{1}{24}  \sin\alpha\, e ^{\phi} F \chi =0 
\label{dilatinouseful} 
\ee
from what we derive
\be
\sin\alpha\, e ^{\phi} F_{ij} \,^j = 2 \partial_i \phi
=4\sin^2\alpha\,\del_i A\ .
\label{traceF}
\ee
where the last equality comes from (\ref{FA1}). 
This is the correct relation between the warp factor and 
dilaton both for type B and type C compactifications 
(see section \ref{s:known}).
In all solutions except type B, the dilaton varies over the
6-dimensional space.

The only equation that we have not used so far is the one we get by
substructing 
(\ref{grav2}) from (\ref{grav1}):
\be
\left(\frac{i}{2} \sin(2\alpha)   \gamma_m\,^n \partial_n A   
-\frac{1}{8} H_m   + \frac{i}{8}\cos\alpha\,  e^{\phi} F_m 
-\frac{i}{2}\del_m\alpha\right)\chi=0
\ .\label{grav4}
\ee
For $m=i$, we end up with
\be\label{grav5}
\left(H-i\cos\alpha e^\phi F\right)_{ij}{}^j = 2i\del_i\alpha-
2i\sin(2\alpha)\del_i A\ ,\ee
and, for $m=\bi$, we get
\be
\left(H-i\cos\alpha e^\phi F\right)_{\bi\bj}{}^{\bj} = 
2i\del_{\bi}\alpha
+2i\sin(2\alpha)\del_{\bi} A\label{grav6}
\ee
and
\be
\left[-\left(H-i\cos\alpha e^\phi F\right)_{\bi jk}\gamma^{jk}
+4i\sin(2\alpha)\del_j A\gamma_{\bi}{}^j\right]\chi =0\ .\label{grav7}
\ee
Taking the complex conjugate of (\ref{grav6}) and comparing with
(\ref{tracesFH2}), we get a relationship between the function $\alpha$
in the spinor ansatz and the warp factor:
\be
\del_i\alpha=-\sin(2\alpha)\del_i A
\label{eqforAalpha}
\ee
while adding (\ref{tracesFH2}) and (\ref{grav5}) we get an equation for
the trace of $H_{(3)}$~:
\be\label{final}
H_{ij}{}^j=2i\del_i\alpha
\ee
Finally,  we get from (\ref{grav7})
\be\label{21HF}
\cos\alpha\, F_{\bi jk} +ie^{-\phi} H_{\bi jk}=-4 e^{-\phi} 
\sin(2\alpha) 
g_{\bi [j} \partial_{k]}A
\ee
The complex conjugate of this equation  for the case $\alpha=0$ is the
type B condition of having no (1,2) piece of $G_{(3)}$. 

From these equations we can see that $\alpha$ being constant gives
nontrivial solutions only when this constant is zero or $\pi/2$, which
corresponds to type B and C. So in the class of solutions with
supersymmetry parameters of the form (\ref{ansatzspinors}), type B
and C are very special ones. 

\subsection{Scalar Relations and Complex Structure}\label{ss:scalarcomplex}

We can collect the information we have so far to write the 4D and  6D
warp factors, the 4-form potential, and the dilaton in terms of
$\alpha$. After we have done this, we will be able to relate the 3-form
fluxes to the complex structure, and finally get $\alpha$ in terms of
the complex structure. 

Let us first obtain $A,A', \phi$ and $h$ in terms of $\alpha$. From
(\ref{eqforAalpha}) we get $A$ in terms of $\alpha$. Then, (\ref{relhA})
gives $h$ as a function of $\alpha$. We use the last equality in
(\ref{traceF}) to get the dilaton in terms of $\alpha$, and finally 
we get $A'$ from (\ref{AA'}). The result is 
\bea
A&=& -\frac{1}{2}\ln\tan\alpha +A_0\label{Aalpha}\\
h&=& e^{4A_0-\phi_0} \cot^2 \alpha +h_0 = e^{4A-\phi_0}+h_0\label{halpha}\\
\phi&=& \ln\cos\alpha +\phi_0\label{phialpha}\\
A^\prime&=& -\frac{1}{2}\ln\sin(2\alpha) +A^\prime_0\ .\label{Apalpha}
\eea
So indeed
the only possible solutions with $\alpha$ constant for our spinor
ansatz are type B and C.
Note that these scalar solutions appear to be singular 
when $\alpha\to 0,\pi/2$, which are the
type B and C limits.  Still, we recover the known solutions by looking
at the relations between $A,h,\phi$ and $A'$ 
implied by (\ref{Aalpha}-\ref{Apalpha}), i.e.:
when $\alpha \rightarrow 0$ (type B), we get 
$A' \sim A$, $h \sim e^{4A-\phi}$, $\phi=\phi_0$, which is what we 
expect for
type B; 
$\alpha \rightarrow \pi/2$ (type C) gives   
$A' \sim -A$, $h=0$, $\phi=2A$, which is what we expect for
type C.  (And these relations are implied by the differential 
equations,
also, even if you disbelieve the solutions above.)

Let us make the type B and C limits more precise.
We can include the type B and C cases if we allow the integration
constants (denoted by subscript $0$) to be infinite, so that they 
cancel the
divergences due to the dependence on $\alpha$.  Let us consider first
the type B limit, where $\alpha(x)=\delta\beta(x)$ with $\delta\to 0$ a 
constant.  Then the scalar equations (\ref{Aalpha},\ref{halpha},
\ref{phialpha},\ref{Apalpha}) become
\bea
A&=& -\frac{1}{2}\ln\beta +\hat A_0, \ A_0 =\hat 
A_0+\frac{1}{2}\ln\delta 
\nonumber\\
h&=& e^{4\hat A_0} e^{-\phi_0} \beta^{-2} +h_0=-e^{-\phi}e^{4A}+h_0
\nonumber\\
\phi&=& \phi_0=\mathnormal{constant}\nonumber\\
A^\prime&=& -\frac{1}{2}\ln\beta +\hat A^\prime_0,\ A^\prime_0 
=\hat{A}^\prime_0+\frac{1}{2}\ln(2\delta),\ A^\prime = A+
\mathnormal{constant}\ .\label{renormB}
\eea
The relations (\ref{renormB}) are indeed what we expect in type B 
solutions
(the constant in the relation between $A^\prime$ and $A$ is usually set 
to
zero).  Additionally, as we briefly noted before, when $\delta\to 0$, eqn (\ref{21HF}) shows that
the $(2,1)$ part of $\b G_3$  vanishes, as 
does
the $(1,2)$ part of $G_3$.  Finally, eqn (\ref{tracesFH2}) 
also shows that $G_3$ is primitive, so we we have the well-known 
result
that $G$ should be a primitive $(2,1)$ form for type B solutions.

For type C, let $\alpha(x^m) = \pi/2 -\delta\beta(x^m)$, where again
$\delta\to 0$ is a constant.  Then the scalars become
\bea
A&=& \frac{1}{2}\ln\beta +\hat{A}_0,\ A_0 
=\hat{A}_0-\frac{1}{2}\ln\delta
\nonumber\\
h&=& e^{4\hat{A}_0}e^{-\hat\phi_0} \delta\beta^2 +h_0\to h_0=
\mathnormal{constant}\nonumber\\
\phi&=& \ln\beta +\hat\phi_0,\ \phi_0=\hat\phi_0-\ln\delta,\ 
\phi=2A+\mathnormal{constant}\nonumber\\
A^\prime &=& -\frac{1}{2}\ln\beta +\hat A^\prime_0,\ A^\prime_0 
=\hat{A}^\prime_0+\frac{1}{2}\ln(2\delta),\ A^\prime = 
-A+\mathnormal{constant}\ .\label{renormC}
\eea
These are, in fact, the expected relations among the scalars for type 
C.
In particular, note that $2A=-2A'=\phi$, as in section \ref{ss:typesAandC}.
Additionally, the NS-NS 3-form vanishes because of 
(\ref{21HF}) and its conjugate (along with the fact that no 
$(0,3),(3,0)$
fluxes are allowed).  This is just what we would expect for a D5-brane
type background.

Solutions that go to the type B or C limit at some position pose an
interesting problem.  In some cases, the divergent scalars give what 
we
expect due to the presence of a source, such as the dilaton in the 
presence
of a D5-brane.  Then we would not want to renormalize out the 
divergence.
We will examine this case in more detail in the example of the D5/D3 
bound
state.
In cases where the solution approaches a limit at infinity,
the situation is much more complicated, depending on the expected 
solution.
For example, if the solution should go to type B AdS at infinity, the
warp factor is expected to diverge in a certain way, so, again, we 
might
not have to renormalize out that divergence.

Let us now get the 3-form fluxes and $\alpha$ in 
terms of the complex structure
and holomorphic (3,0) form of the manifold. This derivation follows
that of Strominger's \cite{Strominger:1986uh}, since our equations
(\ref{cov+torsion}) and
(\ref{dilatinouseful})
have the same form as those that appear in the heterotic case, with 
the NS-NS flux replaced by the R-R flux combined with a function of $\alpha$. 

The Killing spinor equation (\ref{cov+torsion}) implies that the 
supersymmetry parameters feel a torsion in the metric $d\tilde s_6^2$;
the torsion is equal to 
\be\label{torsion}
T=-\frac{1}{2} \sin \alpha\, e^{\phi+2A'} 
F=-\frac{1}{4}e^{\phi_0+2A'_0} F
\ .\ee
Clearly, if we go to the type B limit, the torsion vanishes, which 
means
that $d\tilde s^2_6$ is a Calabi-Yau manifold, which we also expect.  
The following considerations do not apply in that limit because of 
division
by zero problems, but they do apply in all other solutions. Note that
the
only solution without torsion is type B.

First of all, equation (\ref{cov+torsion}) implies that there is an 
almost
complex structure
\be\label{acs}
\tilde J_m{}^n = i\chi^{\prime \dagger} \tilde\gamma_m{}^n\chi^\prime\ 
.\ee
The complex structure is covariantly constant with respect to the
connection with the torsion given in (\ref{torsion}). 
Therefore it is possible to show that $\tilde J$ is an integrable 
complex
structure and satisfies $\tilde J = i\tilde g_{i\bj}dz^i d\b z^{\bj}$.
Additionally, there is an expression for the RR 3-form flux in terms of 
the
fundamental two form, as in Strominger (see subsection \ref{ss:typesAandC} 
for appropriate normalizations in IIB):
\be
e ^{\phi_0+2A'_0} F= -2 i (\bar{\del} - \del) \tilde J
\label{Ffromtorsion}
\ee
Again, in the type B solutions, the left hand side of 
(\ref{Ffromtorsion}) 
vanishes due to the divergence of $A'_0$, so we cannot divide by zero
and use the following results.

With this $F_{(3)}$, let's plug into equation (\ref{21HF}).  Using
(\ref{Aalpha}) we get
\be\label{Hfromtorsion}
H=-2 e ^{-2A'_0} \left(\cos^2\alpha\, d\tilde{J}-
\sin(2\alpha) \tilde{J}\wedge d\alpha  \right)\ .
\ee
To get $\alpha$ in terms of the complex structure and the $(3,0)$ form,
we use equation (\ref{dilatinouseful}). It is a standard procedure 
to multiply this equation to  the
left and to the right with $\tilde{\gamma}_n$ and substract the
resulting equations, to get
\be
\chi^{'\dagger} \left([\tilde{\gamma}_n, \tilde{\gamma} ^m] 
\partial_m \phi
+\frac{1}{12} e ^{\phi_0 +2A'_0} F_{mpq} \{\tilde{\gamma}_n, 
\tilde{\gamma}^{mpq}\}\right) \chi^{'*}=0
\ee
This, as in Strominger's case, leads to the equation
\be
-2\nabla_m \phi + {\tilde J}_m\,^n \nabla_q {\tilde J}_n\, ^q=0
\label{relphiJ}
\ee
where we have used the fact that ${\tilde J}$ is ``$F$-covariantly
constant'', i.e. $\nabla_m {\tilde J}_n{}^p -\frac{1}{4}e^{\phi_0+2A'_0} 
F_{sm}{}^p {\tilde J}_n{}^s -\frac{1}{4}e^{\phi_0+2A'_0} 
F_{mn}{}^s {\tilde J}_s{}^p=0$.

Now construct the holomorphic $(3,0)$ form $\Omega$ as follows
\be
\Omega= e ^{2\phi} \chi^{'\dagger} {\tilde \gamma}_{ijk} \chi^{'*} dz 
^i
dz ^j dz ^k \label{holo30}
\ee
(to be a $(3,0)$ form we need to construct it with the positive chirality
spinor $\chi^{'*}$, which obeys the same two equations
(\ref{cov+torsion}) and (\ref{dilatinouseful}) as $\chi$).

The covariant antiholomorphic derivative gives
\be
\nabla_{\bar l} \Omega_{ijk} = \left(2\nabla_{\bar l} \phi -
\nabla_{\bar l} \phi\right) \Omega_{ijk}
\ee
where the second term
comes from derivatives acting on $\chi^{'*}$. The right hand side of
this, using (\ref{relphiJ}) to write the derivative of the dilaton in
terms of the complex structure, is exactly the difference between the
covariant and the regular antiholomorphic derivatives of $\Omega$. 
This means that $\Omega$ is holomorphic (for more details, see
\cite{Strominger:1986uh}).

The norm of $\Omega$ can be obtained by
\be
\nabla_{\bar l} |\Omega|= 2\nabla_{\bar l} \phi |\Omega|
\ee
where we have used $\nabla_{\bar l} {\bar \Omega}_{\bi \bj \bk}=
3\nabla_{\bar l} \phi {\bar \Omega}_{\bi \bj \bk}$. Then 
\be
|\Omega|=e ^{2\phi+\Omega_0}
\ee
and then
\be
\phi=\frac{1}{2}\left(\ln|\Omega| - \Omega_0\right).  \label{phiomega}\ee
Comparing with (\ref{phialpha}) we get
\be
\cos \alpha=|\Omega|^{1/2} \, , \ \phi_0=-\frac{\Omega_0}{2}.
\label{normomega}\ee
The relationship between $\alpha$ and the norm of the (3,0) form
(\ref{normomega}) is
not valid in type C, where $\alpha=\pi/2$ is constant, but the norm
of $\Omega$ is not. In that case, Eq. (\ref{phiomega}) is still valid,
but we should not use (\ref{phialpha}) to relate $\Omega$ and
$\alpha$.
 
Finally, the dilatino implies then:
\be
d ^{\dagger}{\tilde J} + i ({\bar \partial}-\partial) |\Omega|=0.
\ee

\subsection{Bianchi Identities}\label{ss:bianchis}

Now we turn to Bianchi identities 
\be 
dF=(2\pi)^2\ap \rho_5,\quad dH=0, \quad dF_{(5)}= 
H \wedge F+ (2\pi)^4\ap{}^2 \rho_3\ .
\ee
where $F$ is as before the 3-form RR flux. Note that we do not 
include any NS5-brane sources because they should 
lie 
outside this ansatz; in general, one could add those sources. 
There are some subtleties, however.  As has been known 
\cite{Strominger:1986uh}, anomaly relations require a modified 3-form
Bianchi identity, a fact which has been studied recently and extensively in
\cite{Becker:2003yv}.  In the type IIB string theory, these modifications
arise from $\alpha'$ corrections to the D9-brane (and O9-plane) world-volume
action.  Additionally, complications can arise without D9-branes; the
pure 5-brane systems presented in \cite{Kachru:2002sk} appear to have
too few D5-branes to cancel the charge from the O5-planes.  The solution
appears to be that the space transverse to the 5-brane world-volumes is
a 4-chain rather than a 4-cycle.  The integral over the boundary of the
4-chain allows precisely the right amount of charge 
non-conservation\footnote{We thank M. Schulz for bringing this problem,
and its solution, to our attention.}. It seems that $\alpha'$ corrections
are not necessary to understand D5-brane charge conservation in the
absence of D9-branes.

The Bianchi identity for $F$ gives
\be
dF=e ^{-\phi_0 -2A'_0} \bar{\del}\del \tilde J=(2\pi)^7\ap{}^4 \rho_5
\ ,\label{delbardelJ}
\ee
which is just as we would get following Strominger \cite{Strominger:1986uh}.
The corresponding equation for $H$ gives
\be
dH= 2 \sin(2\alpha) e ^{-2A'_0} \left(d\alpha
\wedge d \tilde J+d \tilde J\wedge d\alpha\right)=0
\ee
which is satisfied automatically because of the wedge product.

The Bianchi identity for $F_{(5)}$
leads to the equation
\bea
e^{-4A+2A'}\left(\tilde \Del^2 h-4\del^{\tilde m}(A+A')
\del_m h\right)&=& 8i e ^{-\phi_0-4A'_0} \cos \alpha \tilde{*}_6 
\left(\cos \alpha \partial \tilde J \wedge \bar{\partial} \tilde{J}
\right.\nonumber\\
&&\left.+\sin \alpha \tilde{J} \wedge (\bar{\partial} -\partial) \tilde{J}
\wedge d\alpha)\right)
+(2\pi)^4\ap{}^2\tilde{*}_6\rho_3\, .
\eea
It is straightforward but unilluminating to plug in for the scalars in terms
of $\alpha$.  We get a rather nonlinear equation for
$\alpha$ that controls the entire geometry.  In the type B limit, it reduces
to equation (\ref{warppoisson}).  

We should note that the Bianchi identities are troubling for the prospects
of supersymmetric compactifications due to the no-go theorem of 
\cite{Giddings:2001yu}. All compactifications of type IIB string theory
with localized sources that satisfy a certain BPS-like inequality are
subject to the no-go theorem.  D5-branes on 2-cycles
satisfy 
the inequality, but O5-planes violate it and avoid the no-go theorem.
If the inequality is satisfied by all the local sources in the 
compactification, then the no-go theorem implies that the inequality must
be saturated, and the solution (if supersymmetric) must be of pure type B.
Therefore, an interpolating type compactification -- that is, one that is
not pure type B -- should have O5-planes (or 9-branes perhaps).
However, it is
hard to see how D3-brane charge, which is present in all but pure type C
solutions, could be conserved in a supersymmetric
way, as the supersymmetries of O3-planes are incompatible with those of
O5-planes.

\subsection{Summary of results for BC interpolating solution}

We summarize here the results obtained so far for solutions that
interpolate between type B and type C, i.e. where 5-brane sources are
pure D5.

The spinor anstaz used was 
\bea
\varepsilon^{1'} = \zeta\otimes \chi &+& \zeta^* \otimes \chi^* \nonumber\\
\varepsilon^{2'} = i e ^{i\alpha} \zeta\otimes \chi &-& i  e ^{-i\alpha} 
\zeta^* \otimes \chi^* .
\eea
where $\alpha=0$ corresponds to type B, and $\alpha=\pi/2$ to type C.

The metric and 5-form flux are of the form
\bea
ds^2=e ^{2A}\eta_{\mu \nu} dx^{\mu}dx^{\nu} + e ^{-2A'} {\tilde ds}_6^2 \nonumber\\
F_{\mu\nu\lambda\rho m}= e^{-4A} \epsilon_{\mu\nu\lambda\rho}\partial_m 
h. 
\eea
where ${\tilde ds}_6^2$ is a metric for a complex manifold for which the
connection $\tilde{\Del}_m 
-\frac{1}{8} \sin\alpha\, e^{\phi+2A'} F_{mnp} \tilde{\gamma} ^{np}$ 
has $SU(3)$ holonomy. The functions $A,A', h$ and the dilaton
$\phi$ are related to the function $\alpha$ in the spinor ansatz by
\bea
A&=& -\frac{1}{2}\ln\tan\alpha +A_0\nonumber\\
h&=& e^{4A_0-\phi_0} \cot^2 \alpha +h_0 = e^{4A-\phi_0}+h_0\nonumber\\
\phi&=& \ln\cos\alpha +\phi_0\nonumber\\
A^\prime&=& -\frac{1}{2}\ln\sin(2\alpha) +A^\prime_0\ .
\eea
The fluxes obey the equations
\bea
H_{ij}{}^j&=&2i\del_i\alpha \nonumber\\
e^\phi F_{ij}{}^j &=& 4\sin\alpha \del_i A\ \nonumber\\
\left(\cos\alpha\, F -i e ^{-\phi} H\right)_{ij}{}^j &=&0\, \nonumber\\
\left(\cos\alpha\, F +ie^{-\phi} H\right)_{\bi jk}&=&-4 e^{-\phi} 
\sin(2\alpha) 
g_{\bi [j} \partial_{k]}A \ .
\eea

The complex structure $\tilde{J}$ and the holomorphic
(3,0)
form $\Omega$ of the manifold with metric $\tilde{ds}_6^2$ obey
\be
d ^{\dagger}{\tilde J} + i ({\bar \partial}-\partial) |\Omega|=0.
\ee
where the norm of $\Omega$ is related to the function $\alpha$ by
\be
\cos \, \alpha= |\Omega| ^{1/2}
\ee

Finally, Bianchi identities turn into equations for $\tilde{J}$ and
$|\Omega|$ of the form
\bea
e ^{-\phi_0 -2A'_0} \bar{\del}\del \tilde J&=&2\kappa_{10}^2 \rho_5
\nonumber \\
e^{-4A+2A'}\left(\tilde \Del^2 h-4\del^{\tilde m}(A+A')
\del_m h\right)&=& 8i e ^{-\phi_0-4A'_0} \cos \alpha \tilde{*}_6 
\left(\cos \alpha \partial \tilde J \wedge \bar{\partial} \tilde{J}
\right.\nonumber\\
&&\left.+\sin \alpha \tilde{J} \wedge (\bar{\partial} -\partial) \tilde{J}
\wedge d\alpha)\right)
+(2\pi)^4\ap{}^2\tilde{*}_6\rho_3
\eea
where we wrote the second equation as an equation for $h$ instead of
$|\Omega|$ for simplicity.

\section{Generalizing the Solution}\label{s:generalize}
\subsection{Inclusion of R-R Scalar}\label{ss:Cneq0}

We know that type B solutions can include a nonzero R-R scalar $C$,
as long as the complex dilaton-axion $\tau$ is holomorphic, and it is 
also evident from the equations of motion that arbitrary constant $C$ is
compatible with type C solutions.  Therefore, we expect that nontrivial
behavior of $C$ is compatible with our spinor ansatz 
(\ref{ansatzspinors}).  Here we discuss the changes to the relations given
in section \ref{ss:interp1} without going into the details of the
derivations.

First, the internal gravitino variation (\ref{grav3}) now gives
\be
\left[\Del_m +\frac{i}{4}\cos\alpha e^\phi\del_m C-\left(
\frac{1}{2}\del_n A\gamma_m{}^n-\frac{1}{4}\cos\alpha e^{\phi-4A}
\del_n h\gamma_m{}^n\right)
-\frac{1}{8} \sin\alpha\, e ^{\phi} F_m
\right] \chi =0\ ,
\label{gravC3}
\ee
so we rescale by $e^{-2A'}$, as before, with
\be
-\del_m A +\frac{1}{2}\cos\alpha\, e^{\phi-4A}\del_m h
= \del_m A^\prime\ .
\label{AA'C} 
\ee
This is the same as we used before, but now equation (\ref{relhA}) will
be modified.  The differential equation for the spinor is now 
\be\label{covtorsionC}
\left(\tilde{\Del}_m +\frac{i}{4}\cos\alpha e^\phi \del_m C
-\frac{1}{8} \sin\alpha\, e ^{\phi+2A^\prime} 
F'_{mnp}
\tilde{\gamma}^{np}\right) \chi^\prime =0\ .\ee
The new feature is the derivative term from $C$; it is a $U(1)$ connection
\cite{Schwarz:1983qr} on the internal manifold.  It is very important now
to see that there is a complex structure, which actually follows exactly
as before.  That is, the almost complex structure $\tilde J$ defined in
(\ref{acs}) is neutral under the $U(1)$, so the proof of its integrability
is unchanged.

Now we turn to the external components of the gravitino equation.  As before,
the $(3,0)$ and $(0,3)$ parts of $F'=F-CH$ vanish, and equation (\ref{FA1}) is
unchanged except for taking $F\to F'$.  The other equation, (\ref{relhA})
becomes
\be
\del_i h = 4e^{4A} e^{-\phi}\cos\alpha\,\del_i A +ie^{4A}\del_i C
\label{relhAC}
\ee
We should note (anticipating that we recover $\del\b\tau =0$
for type B) that this is $\del h = \del e^{4A-\phi}$ in the type B limit, 
which is exactly
what is needed for force cancellation on a D3-brane.  Since D3-branes are
BPS in type B backgrounds, that result is as expected.

The dilatino equations are derived as before.  Equations (\ref{tracesFH2})
and (\ref{traceF}) become respectively
\be
\cos\alpha\, F'_{ij}{}^j -i e ^{-\phi} H_{ij}{}^j =2i\sin\alpha\del_i C\, .
\label{tracesFH2C}
\ee
and
\be
\sin\alpha\, e ^{\phi} F'_{ij} \,^j = 2 \partial_i \phi
-2i\cos\alpha e^\phi \del_i C\ .
\label{traceFC}
\ee
In particular, we see from (\ref{traceFC}) that $\tau$ is holomorphic 
as $\alpha\to 0$.

Finally, we consider the other information that we get from the internal
gravitino variation.  
Equations (\ref{grav5},\ref{grav6}) are unchanged except to take $F\to F'$
because the $C$ term cancels precisely with a
contribution from $h$.
Also, we find 
\be\label{21HFC}
\cos\alpha\, F_{\bi jk} +ie^{-\phi} H_{\bi jk}=-4 e^{-\phi} 
\sin(2\alpha) 
g_{\bi [j} \partial_{k]}A+2i \sin\alpha e^\phi g_{\bi[j}\del_{k]}C\ .
\ee
To go with equations (\ref{AA'C},\ref{relhAC}), 
(\ref{grav5},\ref{grav6},\ref{21HFC}) give two more independent equations,
which we can write using relations for the fluxes as
\bea
\partial_i \alpha = -\sin(2\alpha) \partial_i A + i \sin \alpha\, e
^{\phi} \partial_i C \nonumber\\
\partial_i \phi= i \cos \alpha\, e ^{\phi} \partial_iC +2 \sin^2\alpha\,
\partial_i A\ .\label{scalarsC}
\eea
There are not enough equations to determine all the scalars in terms of a 
single function now.  This fact is not entirely surprising; in type B 
solutions, the holomorphic $\tau$ is after all independent of the warp
factor.  We can, however, determine
\be
\phi=\ln \sin \alpha +2A +\phi_0\ .
\ee

There is a troubling aspect to the fact that all the scalars are not given
by a single function.  Namely, we can no longer prove that the combination
(\ref{AA'C}) is a total derivative.  After manipulation of the various 
scalar relations, it is possible to show that a necessary and sufficient
condition is that
\be\label{Ccondition1}
\cos\alpha e^\phi \left(\b\del-\del\right) C = dB\ee
is a total derivative.  This is true, at least locally, as long as the
left hand side is closed, or
\be\label{Ccondition2}
2\del\b\del C = -d\left(\phi+\ln\cos\alpha\right)\wedge \left(\b\del-\del
\right) C\ .\ee
This condition seems to be rather complicated to solve, but it is easily
demonstrated to be true in the type C ($C\to \mathnormal{constant}$) and
type B ($\alpha\to 0$, $\tau$ holomorphic) limits.  

Finally, we touch on the equations governing the complex structure.
The equations (\ref{Ffromtorsion},\ref{Hfromtorsion}) can be modified simply.
They become
\bea
F' &=& -i\csc\alpha\, e^{-2A'-\phi} \left(\b\del-\del\right)\tilde J
\label{FfromtorsionC}\\
H &=& e^{-2A'}\left( -\frac{1}{2}\cot\alpha\, d\tilde J -4\sin(2\alpha)\, dA
\wedge \tilde J +2i\sin\alpha e^\phi \left(\del-\b\del\right) C\wedge
\tilde J\right)\ .\label{HfromtorsionC}
\eea
These look more complicated because we do not have simple relations among
the scalars, but they reduce to (\ref{Ffromtorsion},\ref{Hfromtorsion})
for constant $C$.
The $(3,0)$ form defined as in (\ref{holo30}) is no longer holomorphic,
however.  To get the holomorphic $(3,0)$ form, we must assume 
(\ref{Ccondition1}); then $\Omega' = e^{-iB/2} \Omega$ is holomorphic.
In that case, it is possible to derive the relations
\be\label{COmega}
\Del_{\bi}\left( |\Omega'|^{-1/2}\Omega'_{j_1 j_2 j_3}\right)
= -i\del_{\bi} B |\Omega'|^{-1/2}\Omega'_{j_1 j_2 j_3}\ ,\ 
\Del_{i}\left( |\Omega'|^{-3/2}\Omega'_{j_1 j_2 j_3}\right)
= -i\del_{i} B |\Omega'|^{-3/2}\Omega'_{j_1 j_2 j_3}\ee
for the R-R scalar.  It would be interesting to see if these results provide
a topological constraint on the behavior of the R-R axion.

\subsection{S-duality and $(p,q)$5-branes}\label{ss:sduality}

Type A solutions, corresponding to NS5-branes, are much more well-studied
than the type C solutions, so it is useful to examine how our interpolating
solutions dualize to solutions that interpolate between type B and type A
solutions.  These solutions correspond to NS5/D3-brane bound states. 
Let us start by noting the action of the $SL(2,\mathbf{Z})$
duality (hats denote the dual variables) (see \cite{Polchinski:1998rr}):
\bea
\hat\tau&=& \frac{a\tau+b}{c\tau+d}\nonumber\\
\hat g^E_{MN}&=& g^E_{MN}\ ,\ g^E_{MN}=e^{-\phi/2}g_{MN}\nonumber\\
\hat F_{(5)} &=& F_{(5)}\nonumber\\
\begin{bmatrix} \hat H_{(3)}\\ \hat F_{(3)}\end{bmatrix} &=& 
\begin{bmatrix} d & c\\b&a\end{bmatrix}
\begin{bmatrix} H_{(3)}\\ F_{(3)}\end{bmatrix}\ .\label{duality}
\eea

To get the pure type A-B interpolating solution, then, start with the
$C=0$ solution of section \ref{ss:interp1} and dualize taking $b=-c=1$.
Then the scalars become
\bea
\hat A &=& A-\frac{\phi}{4} +\frac{\hat\phi}{4} =
-\frac{1}{2}\sin\alpha +\hat A_0 ,\ \hat A_0\equiv A_0-\frac{\phi_0}{2} 
\nonumber\\
\hat h &=& h = e^{4\hat A_0}e^{-\hat\phi_0} \cot^2\alpha+h_0\nonumber\\
\hat\phi&=& -\phi=-\ln\cos\alpha +\hat\phi_0\ ,\ \hat\phi_0\equiv -\phi_0
\nonumber\\
\hat A'&=& -\frac{1}{2} \ln\sin\alpha +\hat A'_0\ ,\ \hat A'_0\equiv
A'_0-\frac{1}{2}\ln 2 +\frac{\phi_0}{2}\ .\label{dualscalars}
\eea
The fluxes obey the equations
\bea
\sin\alpha e^{-\hat\phi}\hat H_{ij}{}^j&=& 2\del_i\hat\phi\nonumber\\
\hat F_{ij}{}^j &=&2i\del_i\alpha\nonumber\\
\left(\cos\alpha\,\hat H+ie^{\hat\phi}\hat F\right)_{ij}{}^j&=& 0
\nonumber\\
\left(\cos\alpha\,\hat H-ie^{\hat\phi}\hat F\right)_{\bi jk}&=&
4\sin(2\alpha) \hat g_{\bi [j}\del_{k]}\left( \hat A-\hat\phi/2\right)\ .
\label{dualforms}
\eea
It is very easy to check that these correspond to a spinor ansatz
$\varepsilon=e^{\hat A /2} \varepsilon'$, 
\bea
\hat\varepsilon^{1'} &=&(e^{-i\alpha/2}-ie^{i\alpha/2})\zeta\otimes\chi +
(e^{i\alpha/2}+ie^{-i\alpha/2})\zeta^*\otimes\chi^*\nonumber\\
\hat\varepsilon^{2'} &=&(e^{-i\alpha/2}+ie^{i\alpha/2})\zeta\otimes\chi +
(e^{i\alpha/2}-ie^{-i\alpha/2})\zeta^*\otimes\chi^*\ .\label{dualansatz}
\eea
Here $\chi$ is covariantly constant with respect to the torsional
connection.

This is what we expect from the $SL(2,\mathbf{R})$
transformation of the superparameter that can be derived from
\cite{Schwarz:1983qr,Hassan:1999bv},
\be\label{susydual}
\hat\varepsilon^1-i\hat\varepsilon^2 =
e^{\hat\phi/8-\phi/8}
\left(\frac{c\tau+d}{|c\tau+d|}\right)^{1/2} \left(\varepsilon^1-i
\varepsilon^2\right)\ee
for $\tau$ purely imaginary.  The dilaton prefactor comes from the 
transformation of the metric.

Naively, we could continue the $SL(2,\mathbf{Z})$ dualities by shifting
$\tau$ to get $(p,q)$5/D3-brane bound states, but those all have nonzero
asymptotic R-R scalar, and they have vanishing $F'=F-CH$ as 
$\alpha\to \pi/2$.
Indeed, the spinors are the same as those for the NS5-brane, so they are not
the $(p,q)$5-branes we want.  Fortunately, the low-energy supergravity has
the full $SL(2,\mathbf{R})$ as a symmetry, and there are $SL(2,\mathbf{R})$
transformations that take NS5-branes to $(p,q)$5-branes with vanishing
asymptotic R-R scalar \cite{Bergshoeff:1996sq,Poppe:1997uz,Lu:1998vh}.
These transformations can similarly be used to generate the $(p,q)$5/D3-brane 
bound state from the NS5/D3-brane bound state given by (\ref{dualansatz}).
We should note that the $(p,q)$5-brane states have a nonconstant R-R
scalar, but it is determined as all other scalars are.

We might think it is possible to generate the solutions of 
section \ref{ss:Cneq0} in this manner.
However, let us note that the $SL(2,\mathbf{R})$ transformations
acting on the solutions of section \ref{ss:interp1}, those with D5-brane
charge and vanishing R-R scalar, can only yield a constant R-R scalar if
they return D5-brane charge.
Therefore, the solutions of section \ref{ss:Cneq0} cannot be reached by
$SL(2,\mathbf{R})$ transformations.  This is because the varying axion in
\ref{ss:Cneq0} can be thought of as back-reaction to 7-branes.  It would
be interesting to study states dual to the solutions with varying axion,
but the duality transformations would be somewhat messy.

\section{Bound state of D3/D5 branes as a particular interpolating 
solution}\label{s:bound}

Breckenridge, Michaud, and Myers \cite{Breckenridge:1997tt} 
and Costa and Papadopoulos \cite{Costa:1998zd}
found the supergravity
solution 
corresponding to a bound state
of D3/D5-branes by applying T-duality to the solution for a $D_4$
brane located at 
an angle with respect to the direction of T-duality.

The supergravity solution they found corresponds to a D5 brane expanded in the 
$(x^0,x^1,x^2,x^3,x,y)$ directions, with 
D3 brane flux spread uniformly in the $x-y$ plane directions. The ratio of
D3 to D5 charge densities, which is the tangent of the angle of
the original D4-brane in the $x-y$ plane, is
\be
\frac{{\tilde q}_3}{q_5}=-\tan\varphi
\ee
where both ${\tilde q}_3$  and $q_5$ are charge densities in the
$(0,1,2,3,x,y)$ volume.

The metric, dilaton, 2-form and 4-form potentials for the
bound state configuration are as follows
\bea
ds^2&=& \frac{1}{\sqrt{H}} \left(-dx^{0} dx^{0}+\sum_{i=1}^3 dx^idx^i 
\right) +\frac{\sqrt{H}}{1+(H-1) \cos^2 \varphi} 
\left\{ dx^2+dy^2 \right. \nonumber\\
&& \left.+ (1+(H-1) \cos^2 \varphi)\left(dr^2+r^2 \left(d\theta^2+sin^2 
\theta (d\phi_1^2
+\sin^2 \phi_2 d\phi_2^2)\right)\right)
\right \} \label{g35}\\
 \nonumber\\
C_{(4)}&=& \mp \mu l^2 sin\varphi \left(\frac{1+\frac{1}{2}(H-1) \cos^2 
\varphi}{1+(H-1) \cos^{\varphi}}\right) 
\sin^2 \theta \cos\phi_1 dy \wedge dx \wedge d\theta \wedge d\phi_2 
\nonumber\\
&& \pm \frac{\sin\varphi}{H}  dx^{0} \wedge dx^{1} \wedge dx^2 \wedge 
dx^3 \label{h35}\\
\nonumber\\
C_{(2)}&=& \pm \mu l^2 \cos\varphi \sin^2\theta \cos\phi_1 d\theta \wedge 
d\phi_2 \label{C235}\\
\nonumber\\
B_{(2)}&=& \frac{(H-1)\cos\varphi \sin\varphi}{1+(H-1) \cos^2\varphi} dx 
\wedge dy \label{B235}\\
\nonumber\\
e^{2\phi}&=& \left(1+(H-1) \cos^2\varphi\right)^{-1} \label{phi35}
\eea
where $H=1+\frac{\mu l}{2 r^2}$, $\mu$ is some dimensionless constant 
proportional to the number of D5 branes, 
$l$ is a length scale determined by the string length and number of
branes,
$l=(4\pi g N)^{1/4} l_s$, and
$r$ is the transverse coordinate distance to the D5-branes.
The $\pm$ signs correspond to the choice of D3/D5 or 
$\overline{\textnormal{D3}}/\overline{\textnormal{D5}}$ 
bound state.

Comparing both metrics, (\ref{g35}) for the bound state one and (\ref{ansatz})
for the interpolating solution, and using (\ref{Aalpha}), 
we get the bound state warp
factor $H$ in terms of the parameter $\alpha$ in the interpolating solutions: 
\be
\sqrt{H} = \tan\alpha\, e^{-2A_0}
\label{Handalpha}
\ee
From the bound state dilaton (\ref{phi35}), comparing it with 
(\ref{phialpha}), we get
another relationship that involves the parameter $\varphi$~:
\be
\frac{1}{1+(H-1) \cos^2\varphi}=\cos^2 \alpha \,e^{2\phi_0}
\label{Handalpha2}
\ee
These two are consistent if we choose
\be
e^{-\phi_0}=\sin \varphi \, , \qquad e^{-2A_0}=\tan\varphi
\label{constants}
\ee
Then, going back to (\ref{Handalpha}), we can get the radial behavior of
$\alpha$ as
\be
\sqrt{H} =\sqrt{1+\frac{\mu l}{2 r^2}}= \tan\alpha \,\tan\varphi
\ee
$\alpha$ has the expected behavior in the limits of small and big $r$, 
i.e.
\be
\alpha \rightarrow_{r \to 0} \pi/2 \, , \qquad   \alpha \rightarrow_{r 
\to \infty} \pi/2-\varphi
\ee
This means that when we are close to the D5 branes, we see the 
D5-brane solution 
(type C, $\alpha=\pi/2$), and when we are far away
we see the D3-brane solution (type B, $\alpha=0$) mixing with the D5, 
with a ``strength'' 
related to $\varphi$. $\varphi \to \pi/2$, which corresponds in the
bound state to no D5-charge, gives a type B solution without 3-form flux.
 
We want to check now that the whole solution for a D3/D5 bound state lies in
the class of solutions we found. Let us start with the 4-form
potential. From the $dx^0 \wedge dx^1 \wedge dx^2\wedge dx^3$ term in
(\ref{h35}), it is easy to extract the function $h$ appearing in our ansatz
(\ref{ansatz})
\be
h=\pm \frac{\sin\varphi}{H}=\pm e^{-\phi_0+4A_0} \cot^2\alpha
\ee
where in the second equality we have used (\ref{Handalpha}) and
(\ref{constants}). The result is in perfect agreement with (\ref{halpha}) 
if we choose $h_0=0$ (and the upper sign). 

To check the agreement in the 3-form fluxes, we need to split the 
6-dimensional metric in (\ref{g35})
in a warp factor and $\tilde{g}_{mn}$. Everything is consistent 
if we consider 
the splitting
\bea
e^{-2A'}&=& \frac{\sqrt{H}}{1+(H-1) \cos^2 \varphi} 
\label{metricsplitting}\\
d\tilde{s}_6^2&=& dx^2+dy^2+
 (1+(H-1) \cos^2 \varphi)\left\{ dr^2+r^2 \left(d\theta^2+\sin^2 \theta 
(d\phi_1^2+\sin^2 \phi_2 d\phi_2^2)\right)\right\} 
\nonumber
\eea
Then, comparing with our form for the 6D warp factor (\ref{Apalpha}),
 and using 
(\ref{Handalpha}-\ref{constants}) we get the constant in $A'$ in terms
 of the parameter $\varphi$ as
\be
e^{2A'_0}=\sin (2\varphi)\ .
\ee

The RR 3-form flux is, according to (\ref{C235})
\be
F_{(3)}=dC_{(2)}=-\mu l^2 \cos\varphi\, \sin^2\theta\, \sin\phi_1 \, 
d\phi_1 \wedge 
d\theta \wedge d\phi_2
\ee
where we have used the upper sign. This would agree with our $F_{(3)}$ 
(\ref{Ffromtorsion}) if
\be
\mu l^2 \cos^2\varphi\, \sin^2\theta\, 
\sin\phi_1 \, d\phi_1 \wedge d\theta \wedge d\phi_2 =  
i(\bar{\del} - \del) \tilde J
\label{Jphi}
\ee
The metric $d\tilde{s}_6^2$ in (\ref{metricsplitting}) splits into a 
flat 2-dimensional metric times a conformally flat 4-dimensional
one. Then, the only nonzero derivatives
of the complex structure $\tilde J$ are 
in these 4 dimensions, and are proportional to $H' \cos^2 
\varphi$ (prime denotes a derivative
with respect to $r$), which is proportional to $\mu l^2 \cos^2 \varphi$. 
After a long but straightforward
calculation that we will not reproduce here, we find precisely (\ref{Jphi}).

The agreement between NS-NS fluxes is easier to check. From 
(\ref{B235}), the NS-NS 3-form flux for the D3/D5 
system is
\be
H_{(3)}= dB_{(2)}= \frac{H'}{(1+(H-1) \cos^2 \varphi)^2} \cos \varphi\, 
\sin \varphi\, dr \wedge dx \wedge dy
\label{theirH3}
\ee
For us, $H_{(3)}$ is (Eq. (\ref{Hfromtorsion}))
\bea
 H_{(3)} &=&-2 e ^{-2A'_0} \left(\cos^2\alpha\, d\tilde{J}-
\sin(2\alpha) \tilde{J}\wedge d\alpha  \right) \nonumber\\
&=& - \frac{\tan \varphi}{1+(H-1) \cos^2 \varphi}d\tilde{J}
+2 \frac{\sqrt{H}}{1+(H-1) \cos^2 \varphi} \tilde{J}\wedge d\alpha 
\label{Hus}
\eea
The first term in this equation involves only the 4-dimensional 
$(r,\theta, \phi_1,\phi_2)$ conformally flat part of the metric, where
\be \label{Jprime}
d\tilde{J}= H' \cos^2\varphi\, dr \wedge \tilde{J}_{4}
\ee
The second term involves the whole complex structure in 6 dimensions 
and a derivative of $\alpha$ which, using
(\ref{Handalpha}) and (\ref{constants}), is
\be \label{alphaprime}
d\alpha=\frac{1}{2 \sqrt{H} (1+(H-1) \cos^2 \varphi)} \cos\varphi\, 
\sin\varphi  H' dr
\ee
Inserting (\ref{Jprime}) and (\ref{alphaprime}) in (\ref{Hus}), we get
\be
H_{(3)}= - \frac{H'}{(1+(H-1) \cos^2 \varphi)^2} \cos\varphi\, \sin\varphi 
(dr \wedge \tilde{J}_{4} - dr \wedge \tilde{J}_{4}) \ ,
\ee
exactly (\ref{theirH3}).

We can conclude that the whole D3/D5 bound state solution of 
\cite{Breckenridge:1997tt,Costa:1998zd}
is in the class of interpolating solutions found. 
This constitutes a nontrivial 
check to our solutions.  

\section{Discussion}\label{s:discuss}

Using our spinor ansatz, we found that there are supersymmetric solutions
of type IIB supergravity that interpolate
between 
type B and type A (or C) solutions. When the axion is zero, we were able to 
write all the functions parametrizing the
solutions (4 and 6D warp factors $A$ and $A'$, the ``potential'' $h$ for 
the 5-form, and dilaton $\phi$),
in terms of a single function $\alpha(x^m)$.
To find the actual form of the solutions, we need to solve
 a nonlinear equation for the
function $\alpha$. These functions of $\alpha$ are smooth for
the interval $\alpha \in (0,\pi/2)$, but they diverge at the
extrema (where it is possible to renormalize away the divergence).  
 We might have anticipated this behavior, from the 
fact that type B and C correspond to special limits of the sources: 
no D3 charge (type C) or D5-branes wrapping a vanishing 2-cycle
(type B). Also, is only at these extremal points in spinor space
that a solution with constant $\alpha$ is possible.

We have also demonstrated how the axion affects the solution; the scalars
are no longer determined in terms of a single function.  We expect this
behavior
from type B systems with D7-branes, in which the complex dilaton-axion
$\tau$ is a holomorphic function independent of the warp factor.  Therefore,
we have generalized D7-brane solutions.  

Our work is closely related to the recent warped M-theory backgrounds 
with G-flux of Martelli and Sparks \cite{Martelli:2003ki}. In this
paper,
the authors generalize the spinor of Becker and Becker's
compactifications to 3 dimensions  \cite{Becker:1996gj} allowing for
8-dimensional spinors with both chiralities. These backgrounds
correspond to space filling M2-branes 
as well as M5-branes. Their background fields depend on a function 
$\zeta$, which is related to the norm of the spinors of each chirality.
Doing a reduction to 10 dimensions and T-dualizing, we get the same
features
of our AB interpolating solutions, if we identify their $\sin \zeta$ with our
$\cos \alpha$. The M2/M5 bound state
solution of \cite{Izquierdo:1996ms} is in the class of
compactifications studied by Martelli and Sparks, while the D3/D5
bound states of \cite{Breckenridge:1997tt,Costa:1998zd} are in our class.

There are two points to make, however.  The first is that we have
used a very particular spinor ansatz, and it is unclear if our solutions
are as general as those of \cite{Martelli:2003ki}, even including
$SL(2,\mathbf{R})$ transformations.  The other is related to the R-R axion
as discussed in section \ref{ss:Cneq0}.  In the type B case, the nontrivial
axion is sourced by D7-branes, which are dual to Kaluza-Klein monopoles
in the M theory.  To be consistent with the results of \cite{Martelli:2003ki},
these monopoles are related to a nontrivial $U(1)$ fibration in the 
manifold of $G_2$ structure.  It would be interesting to see how the relations
(\ref{COmega}) arise in that context.\footnote{We thank D. Martelli and
J. Sparks for discussions of this issue.}

From the AdS/CFT viewpont, it would be interesting to study the gauge
theory duals of these interpolating solutions, in which $\alpha$ regulates
the renormalization flow.  For example, a 
solution could interpolate between the Klebanov-Strassler solution 
\cite{Klebanov:2000hb} in the UV and the Maldacena-Nu{\~n}ez solution
\cite{Maldacena:2000yy} in the IR (or vice-versa).  Such a solution could
provide interesting insights into the relation of the theories on the
branes (especially since the UV theory in \cite{Maldacena:2000yy} is a 
little string theory), as well as providing interesting gauge theory
dynamics.

There are some other known AdS renormalization flows that are related to
our ansatz.  For example, the $\N=2$ flow of 
\cite{Pilch:2000ue,Pilch:2003jg} seems to be a related to our interpolating
solutions with $(p,q)$5-brane charge.  It has been argued 
\cite{Martelli:2003ki} that the $\N=2$ M-theory flow of
\cite{Gowdigere:2003jf} is such an interpolating solution.  Also of
interest is the $\N=1^*$ flow of Polchinski and Strassler 
\cite{Polchinski:2000uf}, which is only known perturbatively.
This solution is a 5-brane/D3-brane bound state, but the perturbative
solution includes $(0,3)$ fluxes in the usual complex coordinates
\cite{Grana:2000jj}.   It is
possible that the complex coordinates should be chosen differently, or
the supersymmetry spinors for $\N=1^*$ may be more complicated than our
ansatz.  Nevertheless, we hope that our work will be a step toward finding
the exact solution.
 
Solutions that are not pure type B are also
interesting for phenomenology. It was shown in \cite{Grana:2002tu} that 
the fields on a D3-brane do not couple to a type B background. 
For standard-like models made out of D3-branes, the only way to get
fermion masses from the fluxes is by embedding the branes on
backgrounds that are not pure type B. 

In conclusion, we have made progress toward finding a unified description 
of all supersymmetric solutions of type IIB supergravity.  More progress
in that direction would be very helpful in finding exact gravity duals
for gauge theories, as we have indicated.

\begin{acknowledgments}
We are grateful to Peter Kaste, Dario Martelli, Ruben Minasian, 
Joseph Polchinski, Michael Schulz, and
Alessandro Tomasiello for very useful discussions. 
In particular, we thank Dario Martelli and James Sparks for discussing
their results \cite{Martelli:2003ki} prior to public release.
The work of A.F. was supported by National Science Foundation grant
PHY00-98395.
The research of M.G. is partially supported by EEC contracts
HPRN-CT-2000-00122, HPRN-CT-2000-00131, HPRN-CT-2000-00148.
\end{acknowledgments}

\bibliography{newinterpolating}

\end{document}